\documentstyle[12pt]{article}
\def\lsim{\mathrel{\rlap{\raise 2.5pt \hbox{$<$}}\lower 2.5pt}}
\def\gsim{\mathrel{\rlap{\raise 2.5pt \hbox{$>$}}\lower 2.5pt}}
\begin{document}
\thispagestyle{empty}
\begin{small}
\begin{flushright}
BUTP-95/39,  hep-ph/9512271\\
\end{flushright}
\end{small}
\vspace{-3mm}
\begin{center}
{\bf{\Large
CP Violation in Heavy Neutrino Mediated $e^- e^- \rightarrow W^- W^-$$^*$}}
\vskip 0.5cm
B. Ananthanarayan\\
P. Minkowski\\

Institut f\"ur Theoretische Physik,\\
 Universit\"at Bern, 5 Sidlerstrasse,\\
CH 3012, Bern, Switzerland.\\
\vskip 3cm
\end{center}
\begin{abstract}
We consider the reaction $e^- e^- \rightarrow W^- W^-$ mediated
by possible heavy neutrino exchange at future LINAC energies of
$\sqrt{s}>> 2 m_W$.  This reaction is sensitive to CP
phases of the neutrino mixing matrices, even at the level of Born
amplitudes.  Certain integrated cross-sections are shown to have
the power to resolve the CP phases when the experimental configurations
are varied.  Asymmetries sensitive to CP violation
(involving initial QED phases) for
$e^- e^-$ and $e^+ e^+$ reactions
are constructed and their consequences
considered.

\end{abstract}
\noindent{\underline{\hspace{11.6cm}}}\\[-2mm]
*{\footnotesize Supported by the Swiss National Science Foundation.}

\newpage

The recent past has witnessed some interest in the construction of
novel LINACs that would collide like sign leptons, viz.,
$e^- e^-,\ e^+ e^+, \ \mu^- \mu^-$ and $\mu^+ \mu^+$ in addition
to conventional $e^+ e^-$ collisions, planned into the next
century at typical center of mass energies of 500 GeV or even
1 TeV~\cite{HC}.  In principle, the novelty of $e^- e^-$ colliders would
include searching for non-resonant phenomena which would be
swamped out at  $e^+ e^-$ colliders and for lepton-number
violating phenomena since the net lepton number of the initial
state would be 2.  In particular, heavy neutrino effects
might mediate reactions of the type:
\begin{equation}
e^-_{L} e^-_{L} \rightarrow W^- W^-.
\end{equation}

In particular, a `minimal' framework within which
the standard model is extended by an arbitrary set of
heavy neutrino flavors (labelled by $\Lambda$)
 with renormalizable couplings
(with 3 light flavors) is the `minimal mass generating
case' (MMC)~\cite{CHPM} wherein reaction (1),
e.g. Fig. 1, is dominated by the
scattering amplitude for the production of longitudinal $W$'s.
This reaction at large $\sqrt{s}$ would be directly
related via crossing to phenomena  such
as neutrinoless double beta decay when quark currents are coupled
to the $W$'s~\cite{dbeta5}.  Furthermore,  CP violating
effects can creep in through the neutrino mass matrix elements
which are in general complex, where not all the phases in the
matrix elements are physically unobservable.  Here we propose that
CP violating effects be looked for in reaction (1) arising from
this source.  It has been shown that the Born amplitude of
Fig. 1 is sufficiently rich so as to exhibit interesting effects
when certain integrated cross-sections are considered as
experimental configurations are varied~\cite{Mink2}.  Furthermore, the
presence of initial state interaction
furnishes a source for the construction of asymmetries between
observables of reaction (1) and those of its CP conjugate
\begin{equation}
e^+ e^+ \rightarrow W^+ W^+.
\end{equation}

We begin with a description of the kinematics of reaction (1)
\begin{equation}
e^-(p_1) e^-(p_2) \rightarrow W^-(p_3) W^-(p_4)
\end{equation}
with
\begin{equation}
s=(p_1+p_2)^2, \ t=(p_1-p_3)^2,\ u=(p_1-p_4)^2
\end{equation}
and with $\sqrt{s}>>m_W$, we have the relation for the cosine
of the c. m. scattering angle:
\begin{equation}
\cos\theta\simeq 1+{2t\over s}\simeq -1+{2u\over s}
\end{equation}
which may be conveniently rewritten as
\begin{equation}
{t\over s}\simeq -\sin^2\theta (\equiv -\zeta), \ {u \over s}
\simeq  -\cos^2\theta (\equiv \zeta-1)
\end{equation}

It has been shown that the amplitude $T$ for the reaction
(1) maybe written as~\cite{CHPM,Lo,GL}
\begin{equation}
T={\sqrt{s}\over v_{ch}^2} \sum_\Lambda
\left( {t\over t-M_\Lambda^2} +{u \over u-M_\Lambda^2}\right)
\end{equation}
where
\begin{equation}
v_{ch}=\left({1\over 2\sqrt{2} G_F}\right)^{1/2}=174.1 {\rm GeV}
\end{equation}
and $G_F$ is
the Fermi constant.

The differential cross-section with respect $\zeta$ defined
in eq. (6) is then
\begin{equation}
{d\sigma \over d\zeta}={1\over 32\pi s} {q_f\over q_i} |T|^2
\end{equation}
and rewriting (7) in terms of (6) and inserting into (9) we obtain
\begin{equation}
{d\sigma \over d\zeta}={1\over 8 \pi v_{ch}^4} |\sum_\Lambda
\xi_\Lambda f(\zeta, x_\Lambda)|^2
\end{equation}
where $x_\Lambda=M_\Lambda^2/s$, $\xi_\Lambda=U_{e\Lambda}M_\Lambda$
 with
\begin{equation}
f(\zeta,x)={1\over 2}\left( {\zeta \over \zeta + x} + {1-\zeta \over
1-\zeta+x}\right)
\end{equation}
\begin{equation}
\sum_\Lambda \xi_\Lambda=0,
\end{equation}
The condition above, eq. (12)
 is a result of the minimal mass generating
case~\cite{PMthui}
which is the framework within which the present effects are discussed.
 In the conventional, see-saw, however, this condition would not
be necessary, but the mixing parameters $U_{e\Lambda}$ would be
small.
Note that in general the $\xi_\Lambda$ are complex and the interesting
effects of possible CP violation reside here.

It is also necessary to introduce a partial wave expansion for
$f(\zeta,x)$:
\begin{equation}
f(\zeta,x)=\sum_{l} (2 l+1) a_l(x) P_l(1-2\zeta)
\end{equation}
where
\begin{equation}
a_l(x)=\delta_{l0}-2\ x\ Q_l(2\ x+1)
\end{equation}
where we have used the standard definition for the Legendre polynomial
of the second kind $Q_{l}(z)$:
\begin{equation}
Q_l(z)={1\over 2}\int_{-1}^1 dy {P_l(y)\over z-y}
\end{equation}

In what follows, we will recall some of the steps already discussed
in the literature~\cite{Mink2}
 and present more complete  details and discussion.
We consider a forward and backward conical section of opening angle
$\theta_0$ characterized by $\Delta$ defined as:
\begin{equation}
\Delta=\cos \theta_0
\end{equation}
We consider the total cross-section in the complement of the
section above:
\begin{equation}
\sigma(\Delta)=\int_{{1-\Delta
\over 2}}^{{1+\Delta \over 2}} d\zeta {d\sigma \over
d\zeta}
\end{equation}

We therefore have the result:
\begin{equation}
\sigma(\Delta)={1\over 8\pi v_{ch}^4}\left[ |\xi_{\Lambda_1}|^2
I(\Delta,x,x)+ {\rm Re}\ (\xi_{\Lambda_1} \xi_{\Lambda_2}^*+
\xi_{\Lambda_2} \xi_{\Lambda_1}^*) I(\Delta,x,y) + |\xi_{\Lambda_2}|^2
I(\Delta,y,y)\right]
\end{equation}
where
\begin{equation}
I(\Delta,x,y)\equiv \int_{{1-\Delta \over 2}}^{{1+\Delta \over 2}}
d\zeta f(\zeta,x) f(\zeta,y)
\end{equation}
and may be expressed in closed form.

Note that in eq. (18) the 2nd term in the square bracket involves
the phases that are now physically observable.

In order to devise a strategy to explore the individual terms
contributing to $\sigma(\Delta)$, it would be necessary to
perform measurements at (a) varying $\Delta$, and/or at (b)
varying $\sqrt{s}$.  We explore the scenarios where $\sqrt{s}$
lies between the masses $M_1$ and $M_2$ and compute $I(\Delta,x,x),
I(\Delta,x,y)$ and $I(\Delta,y,y)$ for a variety of such scenarios
and a variety of $\Delta$.  These results are presented pictorially:
In Fig. 2, we assumed $M_1^2:M_2^2=x:y=1:4$, and present
profiles of the $I$'s as functions of $1/x$, with $1\leq 1/x \leq 4$.
A results of a similar exercises for a ratio $x:y=1:10$ are presented
in Fig. 3  and finally for a ratio $x:y=1:90$ in Fig. 4.
The function $I(\Delta,x,x)$ is bounded from above by $\Delta$.

We finally turn to the issue of detecting CP violation when an
experiment is first performed with $e^-$ beams and then with $e^+$
beams.  Certain integrated cross sections, where the differential
cross-section is weighted with Legendre polynomials $P_L(1-2\zeta)$
provide candidates from which asymmetries may be constructed.
In particular, inserting the partial wave expansion introduced
earlier into the differential cross-section and weight the
later with the Legendre Polynomial above yields the result:
\begin{eqnarray}
& \displaystyle
\sigma_L\equiv {1\over 4}\int_0^1 d\zeta {d\sigma\over d\zeta} P_{L}
(1-2\zeta)= &  \nonumber \\
& \displaystyle  \sum_{l l'} \left ( \begin{array}{ c c c}
l & l' & L \\
0 & 0 & 0 \\
\end{array} \right)^2 (2l+1) (2l'+1) t_l t_{l'}^* &
\end{eqnarray}
where we have used the identity for the
square of the standard Wigner $3\ j$ symbol:
\begin{equation}
{1\over 2} \int_{-1}^{1} dy P_l(y) P_m(y) P_n(y)=
\left( \begin{array}{c c c}
l & m & n \\
0 & 0 & 0 \\
\end{array} \right)^2
\end{equation}
and
\begin{equation}
t_l=\sum_\Lambda \xi_\Lambda a_l(x_\Lambda)=|t_l| e^{i\phi_l}
\end{equation}
Note that we may define an analogous cross-section when the experiment
is performed with right handed positron beams $\overline{\sigma}_L$.

Up until now we have not utilized the presence of the initial state
phase due to the Coulomb interaction between the initial state $e^-$
(and the $e^+$ when the experiment is performed with positron beams)
which results in the phase shift:
\begin{equation}
e^{2i\delta_l^C}={\Gamma(l+1-i\alpha)\over \Gamma(l+1+i
\alpha)}
\end{equation}
For small $\alpha$, this expression has the Taylor expansion:
\begin{equation}
1-2i\alpha\psi(l+1),
\end{equation}
where $\psi(z)$ is the di-gamma function.
Furthermore, the Wigner 3 j symbol implies the triangular rule among
the $l,\ m,\ n$ and vanishes unless this is satisfied.
This implies a selection among the $l$ and $l'$ for given $L$.  Thus,
upon including the initial state Coulomb phase, we may consider the
following difference:
\begin{eqnarray}
& \sigma_L^C-\overline{\sigma}_L^C= & \nonumber \\
& \displaystyle  \sum_{l l'} \left ( \begin{array}{ c c c}
l & l' & L \\
0 & 0 & 0 \\
\end{array} \right)^2 (2l+1) (2l'+1) |t_l t_{l'}|
\left[-2\sin(\phi_l-\phi_{l'})\sin(\delta_l^C-\delta_{l'}^C) \right]&
\end{eqnarray}

In particular, for $L=2$ and
assuming the dominace of the two lowest
(S- and D-) waves, we have the result:

\begin{equation}
\sigma_2^C-\overline{\sigma}_2^C\simeq
-6 \alpha |t_2 t_{0}| \sin(\phi_0-\phi_2)
\end{equation}

\newpage

The resulting asymmetry retaining S- and D- waves is:
\begin{eqnarray}
& \displaystyle {\sigma_2^C-\overline{\sigma}_2^C \over
\frac{1}{2} (\sigma_2^C+\overline{\sigma}_2^C)}= & \nonumber \\
& \displaystyle {-3\alpha \sin (\phi_2-\phi_0) \over
\cos (\phi_2-\phi_0) +|\frac{5t_2}{7t_0}|}
{}.
&
\end{eqnarray}
For a natural range of parameters, such as
$\sin(\phi_2-\phi_0)\sim 0.2-0.3$, this
asymmetry can be at the level of a percent.
This constitutes perhaps the most important result of
this work.

We also note here that it might be interesting to look for such
effects when production of photons in the
initial state occurs.  Such a proposal has already been made in the context
of quark-quark interactions~\cite{NP}.
Also of interest would be to consider the possibility of
final state hard bremsstrahlung.
A final interesting possibility which is not ruled out is one
where the final state $W's$ are strongly interacting.
Such a strongly interacting final state could lead to the enhancement
of the CP violating effects due to large final state phases.

\bigskip

\noindent {\bf Acknowledgement:}  We thank M. Nowakowski for
discussions.

\newpage

\noindent{\Large{\bf Figure Captions}}

\bigskip

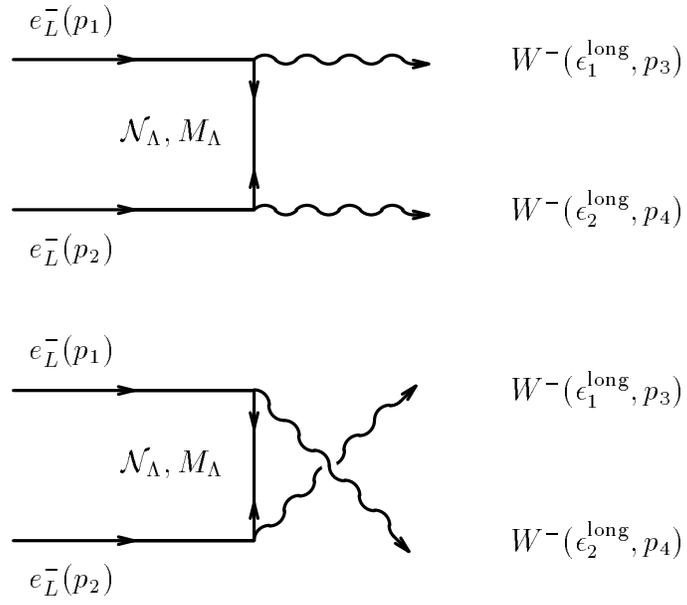
\noindent {\bf Fig. 1.} Graphs for the amplitude T.
Here ${\cal N}_\Lambda$ labels the heavy neutrino
flavor of mass $M_\Lambda$.

\bigskip

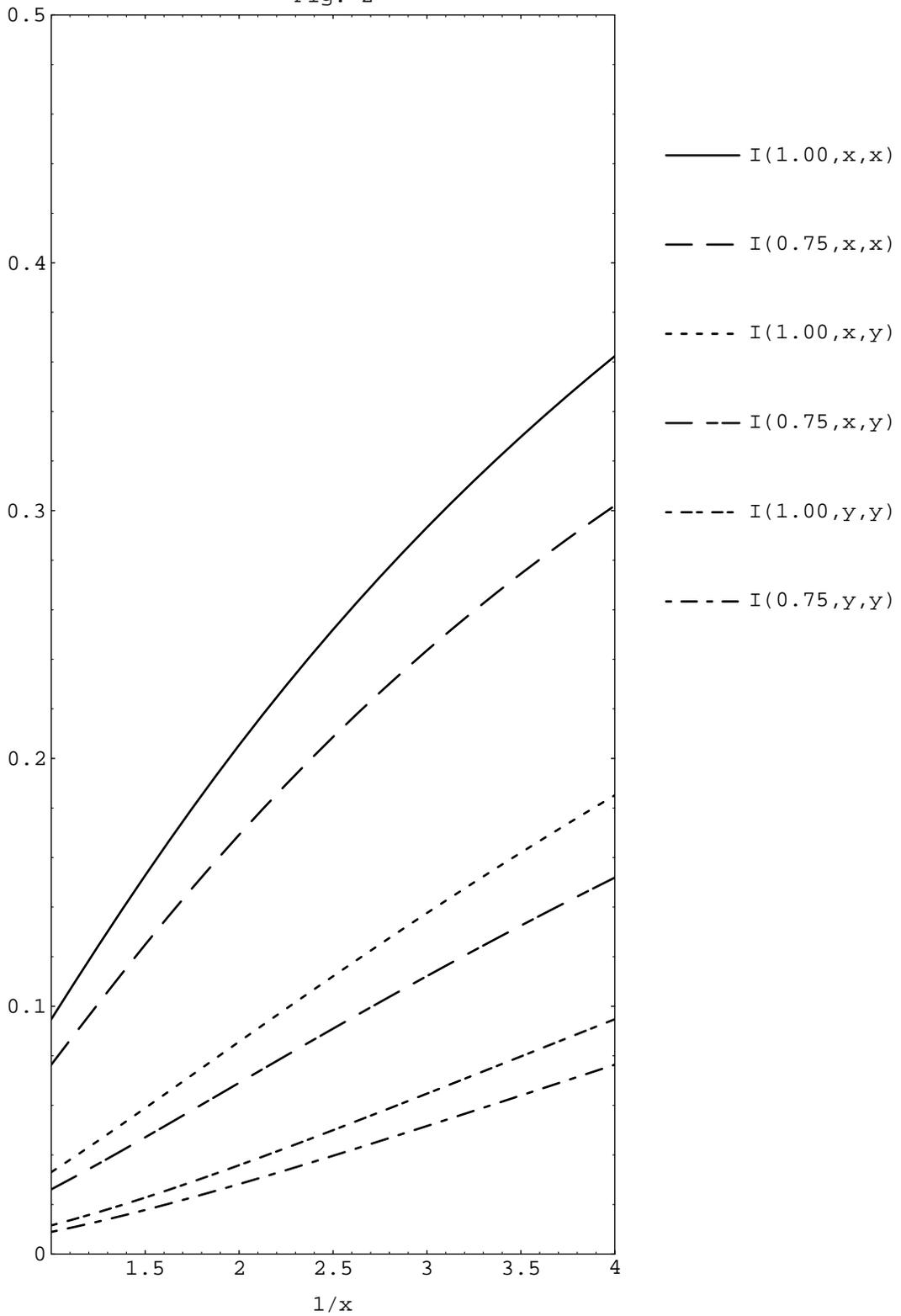
\noindent{\bf Fig. 2.} The integrals $I(\Delta,x,x),
I(\Delta,x,y), I(\Delta,y,y)$ with
$x:y=1:4$ as a function of $1/x$, $1\leq 1/x \leq 4$
for $\Delta=1.00,\ 0.75$.

\bigskip

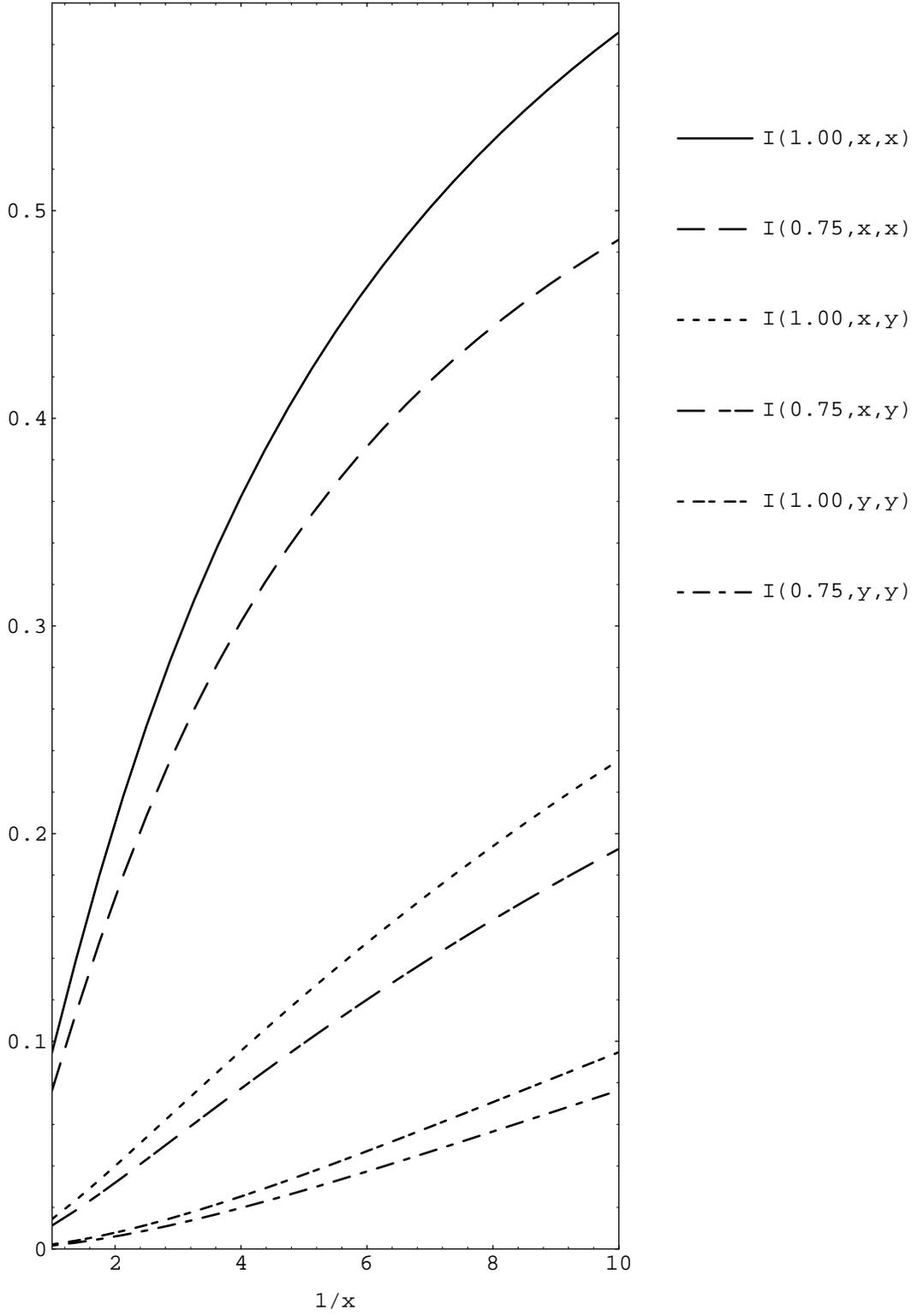
\noindent{\bf Fig. 3.} As in Fig. 2 with $x:y=1:10$ and
$1\leq 1/x \leq 10$.

\bigskip

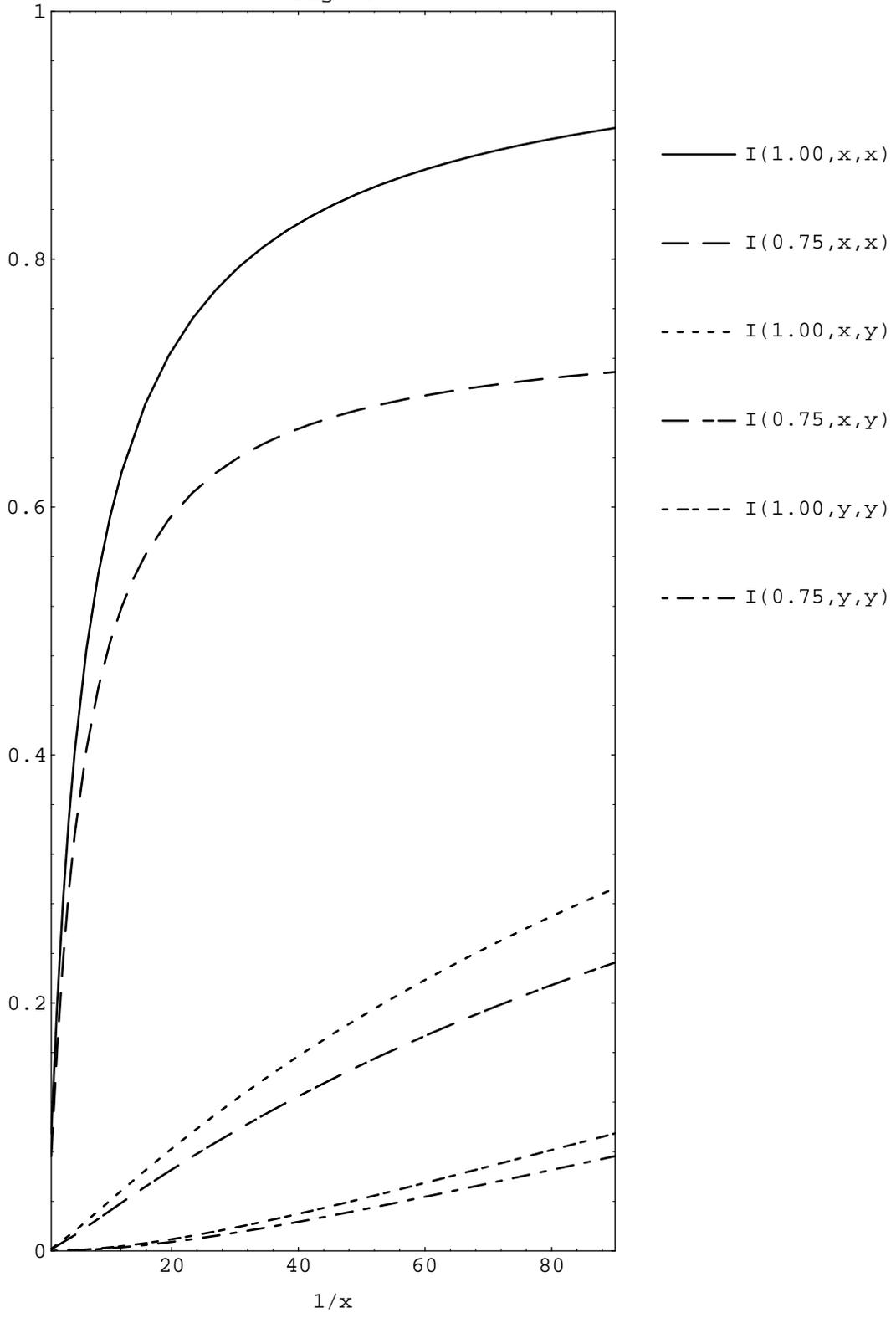
\noindent{\bf Fig. 4.} As in Fig. 2 with $x:y=1:90$ and
$1\leq 1/x \leq 90$.

\end{document}